\documentclass[fleqn]{article}
\usepackage{txfonts}
\usepackage{flafter}
\usepackage{yfonts}
\usepackage{graphicx}
\usepackage{ifthen}
\usepackage[center]{caption}
\newcommand\doingARLO[2][]{%
  \ifx\mmref\undefined #1\else #2\fi
}

\title
      {Extended Emission of Short Gamma-Ray Bursts}

\author{Lin Lin$^1$ \thanks{E-mail:
linlin06@mails.tsinghua.edu.cn},\
  En-Wei Liang$^2$,\
  Bin-Bin Zhang$^3$,\
  Shuang Nan Zhang$^1$\\
  \\
$^1$Department of Physics and Center for Astrophysics, Tsinghua
University, Beijing 100084, China,\\
$^2$Department of Physics, Guangxi University, Nanning 530004,
  China,\\
$^3$Department of Physics and Astronomy, University of Nevada, Las
Vegas, NV 89154, USA}
\begin{document}

\maketitle

\begin{abstract}
Preliminary results of our analysis on the extended emission of
short/medium duration GRBs observed with Swift/BAT are presented.
The Bayesian blocks algorithm is used to analyze the burst durations
and the temporal structure of the lightcurves in different energy
bands. We show here the results of three bursts (GRBs 050724, 061006
and 070714B) that have a prominent soft extended emission component
in our sample. The extended emission of these bursts is a
continuous, flickering-liked component, lasting $\sim 100$ seconds
post the GRB trigger at 15-25 keV bands. Without considering this
component, the three bursts are classified as short GRBs, with
$T_{90}=2\sim 3$ seconds. GRB 060614 has an emission component
similar to the extended emission, but this component has pulse-liked
structure, possibly indicating that this emission component is
different from that observed in GRBs 050724, 061006, and 070714B.
Further analysis on the spectral evolution behavior of the extended
emission component is on going.
\end{abstract}

\begin{description}
\item[Keywords:] Gamma-ray bursts
\item[PACS:] ${98.70.}$Rz
\end{description}

\section*{Introduction}
The \textit{Swift} satellite now is leading a new era in the short
GRB study, and rapid progress has been made. The precise
localization led to redshift measurement and host galaxy connection
for this class of GRBs, favoring mergers of compact object binaries
as the progenitors of the short GRBs \cite{narka07}\cite{zhang07b}.
However, a soft, extended emission component following early hard
spikes observed in the light curves of some short GRBs challenges
the models and blurs the clear picture of long-short classification
of GRBs with a division line at $T_{90}=2$ seconds. The extended
emission lasts tens of seconds\cite{zhang07a}. It was observed by
CGRO/BATSE \cite{norris06} and HETE-2 \cite{donaghy06}. We present a
systematic analysis of this emission component with Swift/BAT
observation from both spectral and temporal features, focusing on
the issue if the extended emission is a distinguished component from
the hard spikes. The Bayesian blocks (BB) method \cite{scargle98} is
used to analyze the BAT light curves and to calculate the duration
of the burst in different energy bands. The time-resolved spectral
analysis for this emission component is also on going.

\section*{Data and Method}
The Swift/BAT data are downloaded from the NASA Swift Archive. We
deduce the BAT data with our code by using the standard BAT
tools\footnote{The details of our data reduction please refer to our
web site at UNLV http://grb.physics.unlv.edu/}. Since the energy
band of BAT is 15-150 keV, we derive the BAT lightcurve in four
energy bands, i.e., 15-25 keV 25-50 keV, 50-100 keV and 100-150 keV,
and fit the observed spectrum with a simple power
${(\emph{F}\propto\nu^{-\Gamma})}$.

We search for the soft, extended emission component based on both
the observed lightcurves and the spectra. We first pick up
candidates based on their lightcurves in different energy bands by
eye, and then analyze the structure measure the burst duration in
different energy bands with the Bayesian block algorithm
\cite{scargle98}.

The BB method is a time-domain algorithm for detecting temporal
structures of a pulse, revealing pulse shapes, and generally
characterizing intensity variations. It rebuilds the raw counting
data into time intervals during which the photon arrival rate is a
perceptible constant. A change point of count rate is obtained by
the Bayesian statistics, and a simple iterative procedure is used to
generate a light curve. Shown in Figure 1 is a test for extracting a
signal that is contaminated with Poisson noise by this method.

\begin{figure}
\includegraphics[width=\textwidth]{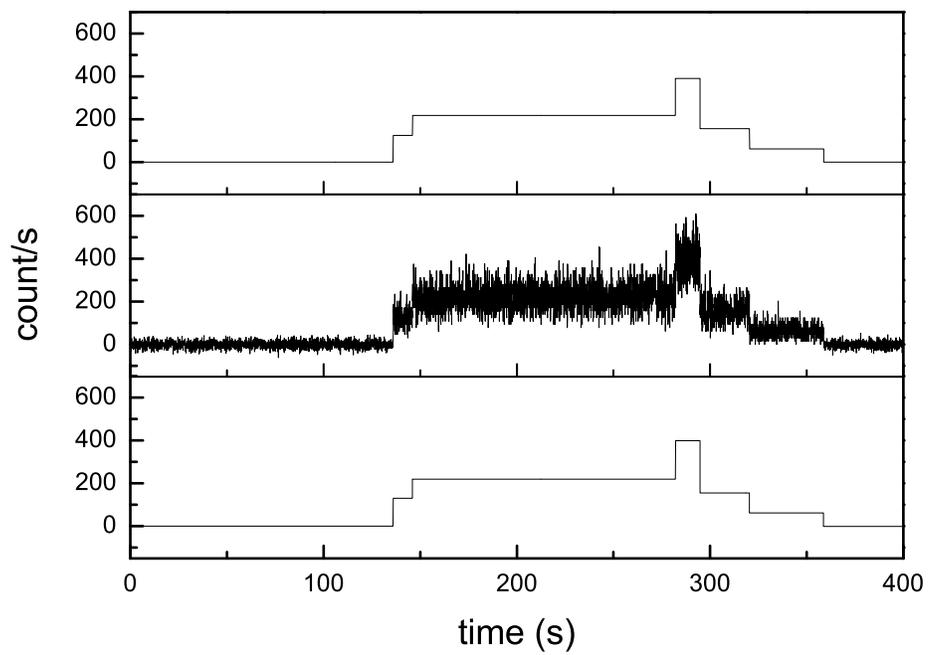}
\caption{Test for extracting a signal that is contaminated by
Poisson noise with the BB method: {\em Top}:  original signal; .
{\em Middle}: signal contaminated with Poisson noise. {\em  Bottom
}: the extracted signal.}
\end{figure}

\section*{Results of Temporal Analysis with the BB Method}
We search for the candidates with the BB method, and a sample of 50
GRBs is obtained. Among the bursts in the sample, GRBs 050724,
061006 and 070714B are the most prominent cases. We present our
preliminary results for these cases in this section. We show the
light curves of GRB 050724 in the four energy bands as an example in
figure 2. The extended emission component is observed in the 15-25
keV band, and its duration become shorter at higher energy bands. It
was not detected at the energy band of 100-150 keV. We calculate the
$T_{90}$ without/with considering the extended emission component,
and the results are shown in Table 1. The extended emission of these
bursts are continuously flickering-like. Considering only the hard
peaks, these three bursts are classified as short GRBs. The soft,
extended emission component is also observed in GRBs
060614\cite{zhang07a} and 070220. It is bright, and also detected in
the energy band of 100-150 keV, as shown in Figure 3. It is
difficult to define a rigorous criterion to separate this component
from the main spikes at early time. Zhang et al. (2007) argued that
time-resolved spectral analysis is an approach to identify it. With
the BB method, we find that this component is pulse-like,
significantly different from that observed in GRBs 050724, 061006,
and 070714B.
\begin{figure}
  \includegraphics[width=\textwidth]{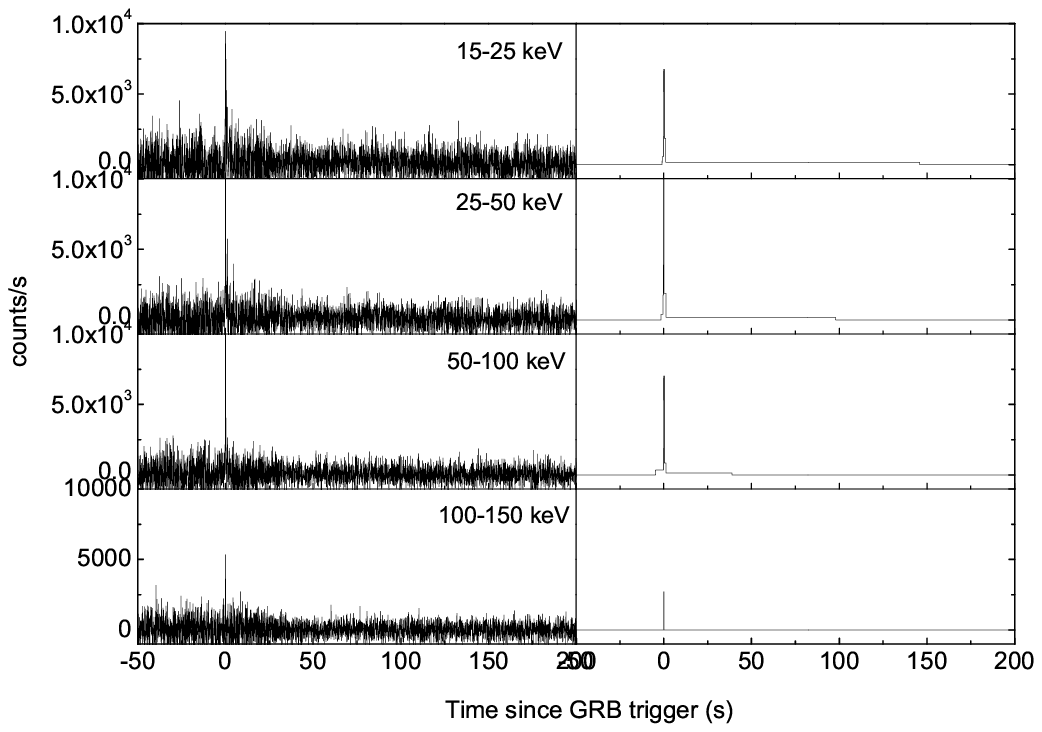}
  \caption{The extended emission component observed in GRB 050724: {\em Left}: original lightcurves; {\em right:}
  subtracted lightcurves.}
\end{figure}

\begin{table*}
 \centering
 \begin{minipage}{140mm}
\caption{$T_{90}$ ${(s)}$ in different energy bands ${(keV)}$.}
\begin{tabular}{*9{c}}
 \hline
 GRB &15-25\footnote{column $2-5$ are $T_{90}$ ${(s)}$ of the whole burst}  &25-50 &50-100 &100-150
 &15-25\footnote{column $6-9$ are ${T_{90}}$ $(s)$ of peak only}
&25-50 &50-100 &100-150 \\
 \hline
050724 & 136.96 & 91.39 & 38.14 & 0.20 & 1.34 & 1.28 & 1.09 & 0.20 \\
061006 & 108.10 & 133.67 & 51.46 & 0.58 & 2.49 & 2.36 & 0.77 & 0.39 \\
070714B & 112.19 & 61.25 & 2.3 & 2.3 & 2.56 & 2.43 & 2.3 & 2.3 \\
060614 & 104.83 & 95.81 & 95.36 & 86.27 & & & & \\
070220 & 72.52 & 106.69 & 63.23 & 37.18 & & & & \\
\hline
\end{tabular}
\end{minipage}
\end{table*}
%\source{Central Statistical Office, UK}

%Similar lightcurve behavior is also observed in GRB 060904B
%(figure 4) and GRB 070704 has a bump hundreds of seconds after the
%main peak in band 1 and 2. If it is true that the extended
%emission has different origin, then the different appearances of
%extended emission imply different properties of the progenitor and
%environment.
%

\begin{figure}
  \includegraphics[width=\textwidth]{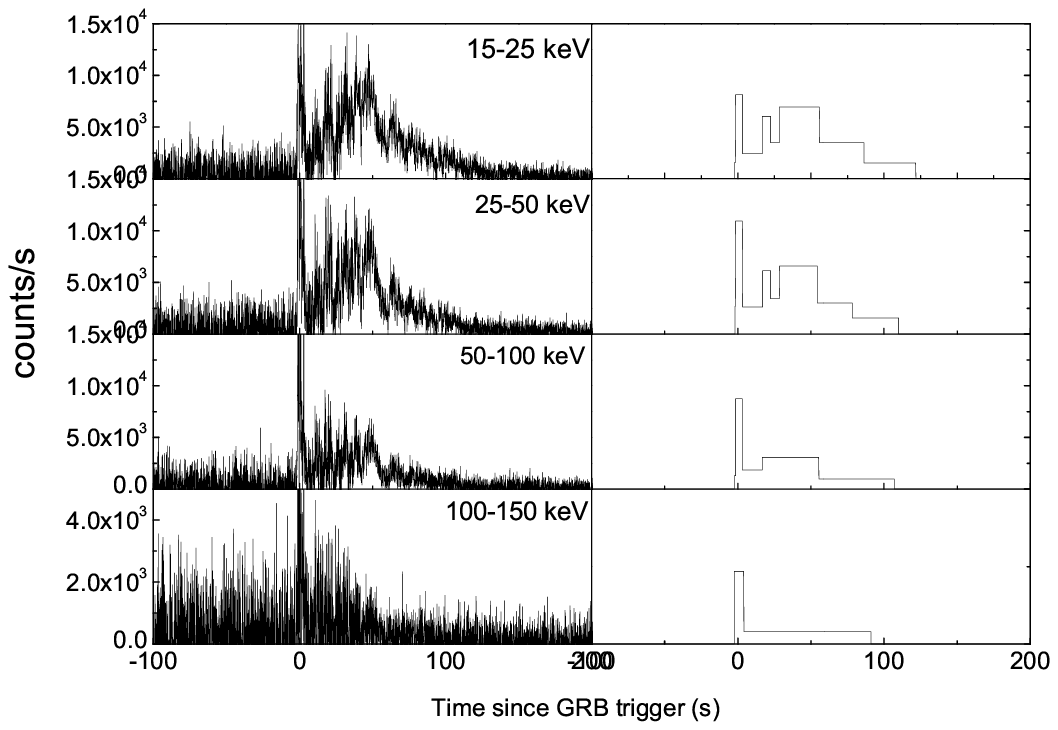}
  \caption{The extended emission component observed in GRB 060614: {\em Left}: original lightcurves; {\em right}
  subtracted lightcurves.}
\end{figure}

\section*{Conclusions}
We have searched for the extended emission of the short GRBs with
the BB method. We first search for the candidates with the BB
method, and a sample of 50 GRBs is obtained. Among the bursts in the
sample, GRBs 050724, 061006 and 070714B are the most prominent
cases. The extended emission from these bursts is a continuous,
flickering-liked component, and is observed only in the low energy
bands. We calculate the $T_{90}$ without/with considering the
extended emission component. Without Considering the extended
emission component the three bursts are classified as short GRBs.

%The duration of the spike get longer in softer band.
%However when considering extended emission, the durations in band
%2 of GRB 050724 and GRB 061006 are longer than those in band 1.
%Thus, these two kinds of T90 change differently with energy bands.
%It implies that the spike and the extended emission have different
%origins.

%\begin{theacknowledgments}
\section*{Acknowledgments} We appreciate Bing Zhang for helpful
discussion. This work was supported by the National Natural Science
Foundation of China under grants No. 10521001, 10733010, 10725313
and 10463001, and by the National Basic Research Program (``973''
Program) of China (Grant 2009CB824800), the research foundation of
Guangxi University ( for EWL), and NASA NNG05GB67G, NNX07AJ64G,
NNX08AN24G (BZ). This project was also in part supported by the
Ministry of Education of China, Directional Research Project of the
Academy of Sciences under project KJCX2-YW-T03.
%\end{theacknowledgments}

% choose bibtex style depending on layout style and options used in
% sample:


\begin{thebibliography}{99}

\bibitem[1]{berger05} Berger, E., et al., 2005, Nature, 438, 988B

\bibitem[2]{donaghy06} Donaghy, T. Q., et al., 2006, astro-ph/0605570v2

\bibitem[3]{krimm05} Krimm, H., et al., 2005, GCN Circular 3667, http://gcn.gsfc.nasa.gov/gcn/gcn3/3667.gcn3 (2005).

\bibitem[4]{krimm06} Krimm, H., et al., 2006, GCN Circular 5704, http://gcn.gsfc.nasa.gov/other/061006.gcn3 (2006).

\bibitem[5]{narka07} Narka, E., 2007, PhR, 442, 166

\bibitem[6]{norris06} Norris, J. P., Bonnell, J. t., 2006, Astrophys. J., 643, 266

\bibitem[7]{scargle98} Scargle, J. D., 1998, Astrophys. J., 504, 405

\bibitem[8]{vaughan06} Vaughan, S., et al., Astrophys. J., 639, 323

\bibitem[9]{zhang06} Zhang, B., 2006, Nature, 444, 1010

\bibitem[10]{zhang07a} Zhang, B., et al., 2007, Astrophys. J., 655, L25

\bibitem[11]{zhang07b} Zhang, B., 2007, Chin. J. Astron. Astrophys., 7, 1
\end{thebibliography}
\end{document}